\documentclass[twoside,11pt]{article}

\usepackage[accepted]{melba}

\usepackage{mwe} 

\usepackage{amsmath,amsfonts}
\usepackage{multirow}
\usepackage{booktabs}
\usepackage{graphicx}
\usepackage{caption}
\usepackage{subcaption}
\usepackage{soul,color}

\melbaid{2023:008}  
\melbaauthors{Qi, Foran, Nosher, and Hacihaliloglu}
\volume{2}
\firstpageno{236}  
\year{2023}  
\datesubmitted{08/2022}  
\datepublished{04/2023}  

\ShortHeadings{Multi-Scale Parallel-Attention Fusion Network}{Qi, Foran, Nosher, and Hacihaliloglu}

\title{Multi-Scale Feature Fusion using Parallel-Attention Block for COVID-19 Chest X-ray Diagnosis}

\author{\name Xiao Qi\orcid{0000-0001-5316-9473} \email xiao.qi@rutgers.edu \\  
	\addr Department of Electrical and Computer Engineering, Rutgers Univeristy, New Brunswick, NJ, USA
	\AND
	\name David J. Foran \email foran@cinj.rutgers.edu\\
	\addr Rutgers Cancer Institute of New Jersey, New Brunswick, NJ, USA
	\AND
	\name John L. Nosher \email nosher@rwjms.rutgers.edu\\
	\addr Department of Radiology, Rutgers Robert Wood Johnson Medical School, New Brunswick, NJ, USA
	\AND
	\name Ilker Hacihaliloglu\orcid{0000-0003-3232-8193} \email ilker.hacihaliloglu@ubc.ca\\
	\addr Department of Radiology, The University of British Columbia, BC, Canada\\
	\addr Department of Medicine, The University of British Columbia, BC, Canada
}

\begin{document}

\maketitle

\begin{abstract}
	Under the global COVID-19 crisis, accurate diagnosis of COVID-19 from Chest X-ray (CXR) images is critical. To reduce intra- and inter-observer variability, during the radiological assessment, computer-aided diagnostic tools have been utilized to supplement medical decision-making and subsequent disease management. Computational methods with high accuracy and robustness are required for rapid triaging of patients and aiding radiologists in the interpretation of the collected data. In this study, we propose a novel multi-feature fusion network using parallel attention blocks to fuse the original CXR images and local-phase feature-enhanced CXR images at multi-scales. We examine our model on various COVID-19 datasets acquired from different organizations to assess the generalization ability. Our experiments demonstrate that our method achieves state-of-art performance and has improved generalization capability, which is crucial for widespread deployment. Our code is available at~\url{https://github.com/endiqq/Multi-Scale-Feature-Fusion}.
\end{abstract}


\section{Introduction}
The novel coronavirus disease 2019 (COVID-19) is an infectious disease caused by severe acute respiratory syndrome coronavirus 2 (SARS-CoV-2). As of August 2022, more than 580 million cases have been reported worldwide (\cite{johns2020covid}). Early and accurate screening of the infected population and isolation from the public is an effective way to prevent and halt the spreading of the virus. Although the initial infection rates are reduced with the newly developed vaccines, the COVID-19 pandemic continues to affect lives around the world. Currently, the real-time polymerase chain reaction (RT-PCR) is considered as the gold standard in the diagnosis of COVID-19 (\cite{rtpcr}). When patients with confirmed COVID-19 arrive at hospitals, initial patient assessment often includes imaging studies such as radiologic examinations, which can be useful for rapid screening of disease (\cite{cxr_exam}). Easily accessible chest X-ray (CXR) imaging has emerged as one of the preferred modalities due to its wide availability, less radiation exposure, and faster image acquisition times. However, the interpretation of CXR images is challenging due to low image resolution and COVID-19 image features being similar to regular pneumonia. Computer-aided diagnosis via deep learning has been investigated to help mitigate these problems and help clinicians during the decision-making process. The novel deep learning methods developed for improved diagnosis of COVID-19 have also led to advancements for similar tasks such as accurately detecting the presence of multiple diseases from CXRs or lung disease identification from ultrasound data (\cite{us_lung, med_img_class}).\\   
The most investigated computational approaches for detection of COVID-19 from CXR scans has been using deep convolutional neural networks (CNNs) with supervised learning (\cite{Wang2020_covid, ozturkcovid2020, covid_tf}). \cite{ozturkcovid2020} proposed a CNN architecture with 17 layers named DarkCovidNet for detection of COVID-19 using CXR iamges. They achieved an average 87.02\% three class classification accuracy using five-fold cross-validation. The method was evaluated on 127 COVID-19, 500 healthy and 500 pneumonia CXR scans. \cite{Wang2020_covid} built a public dataset named COVIDx, which is comprised of a total of 13975 CXR images from 13870 patient case and developed COVID-Net. Their dataset had 358 COVID-19 images obtained from 266 patients. Their model achieved 93.3\% overall accuracy for multiclass classification (Normal vs. Pneumonia vs. COVID-19). \cite{covid_tf} developed an deep transfer learning-based approach for the detection of COVID-19 infection on CXR images using DenseNet201. The proposed model was able to achieve an accuracy of 94\% in detecting COVID-19 and an overall accuracy of 92.19\%. The model was able to achieve an AUC of 0.99 for COVID-19, 0.97 for normal, and 0.97 for pneumonia.\\
Several groups have also investigated various image enhancement methods for improving the representation of CXR image data for improving the diagnostic performance of CNN methods. \cite{covid_enh} investigated five different image enhancement techniques. Among them, the gamma correction-based enhancement technique outperformed other techniques in detecting COVID-19 from the plain and the segmented lung CXR images. The accuracy, precision, sensitivity, F1-score, and specificity were 95.11\%, 94.55\%, 94.56\%, 94.53\%, and 95.59\% respectively were achieved tested on 724 COVID-19, 1771 normal, and 1203 Non-COVID segmented CXR images. \cite{qisl2020} investigated how local phase CXR feature-based image enhancement improves the accuracy of CNN architectures for COVID-19 diagnosis. Three different CXR local phase image features are combined as a multi-feature CXR image. Then a multi-feature CNN with various fusion techniques was proposed for processing both the original CXR image and the multi-feature CXR image. The multi-feature ResNet50 with late-fusion using sum operation achieves 95.57\% and 94.44\% average accuracy tested on two datasets including 3323 COVID-19, 8851 normal, and 6045 pneumonia CXR images.\\
Transformer, a deep neural network based on a self-attention mechanism, was first introduced in the field of natural language processing (NLP) and achieved dominant performance. Inspired by the success of the self-attention mechanism in NLP, Vision Transformer (ViT) has been introduced in 2021 and beats some state-of-the-art (SOTA) convolutional neural networks (CNNs) in some image recognition tasks (\cite{vit}). Recently, ViT has been applied for COVID-19 detection instead of CNN architectures (\cite{chxpert_vit, covid_vit}). \cite{covid_vit} proposed a ViT for COVID-19 detection using CXR images. The authors collected data from three open-source datasets of CXR images including COVID-19, Pneumonia, and Normal cases from multiple datasets to obtain a total of 30,000 images. The proposed ViT model achieved an ACC of 92\% and an AUC score of 0.98 for the multi-class classification (COVID-19, Normal, and Pneumonia). \cite{chxpert_vit} proposed a hybrid model by conjugating the conventional CNN backbone for producing initial feature embedding. Their backbone network was DenseNet121 and trained using a large public Chexpert\cite{chexpert} datasets to obtain the abnormal features. Then, the embedded features from the backbone network were used as the corpus for ViT training. They achieved an average AUC of 0.94, 0.90, and 0.91 on the three datasets including 94 COVID-19, 1020 normal, and 491 other infections.\\
Motivated by the effectiveness of the local phase-based image enhancement technique, CNN and ViT in the detection of COVID-19 from CXR images, we propose a novel multi-scale feature fusion deep learning model. Multi-scale feature fusion has a long history in detection and recognition of objects (\cite{multi_fusion_1, multi_fusion_2}). We first utilize the local phase-based image enhancement technique to generate enhanced CXR images to provide more details of CXR images. Then, a Parallel-Attention (PA) block to encode feature correlation between original CXR images and local-phase enhanced CXR images is developed. Our proposed multi-feature fusion modal has two separate branches for processing the original CXR image and local-phase enhanced CXR image, and PA blocks are deployed after each feature extractor to fuse feature representations from two branches at multi-scales. In the end, the embedded features from two branches are fused together through cross-attention operations. In this work, our contributions and findings include the following: 1) we develop a novel multi-feature fusion deep learning model by deploying the PA blocks at multi-scales. 2) we show that our proposed method obtains improved results compared to state-of-the-art (SOTA) CNNs, ViTs, and multi-feature fusion models. 3) we demonstrate that local-phase feature enhanced CXR-image can further improve diagnostic performance by increasing generalization ability. 4) we also provide clinically interpretable visualization results of the model for helping COVID-19 diagnosis and localization.

\section{Methods}
\subsection{Image Enhancement}

The local phase-based image enhancement method extracts the phase signatures from the image and are intensity invariant (\cite{alessandrini2012myocardial, li2018hybrid}). Therefore, the enhancement results and subsequent image analysis methods are not affected by the intensity variations due to patient characteristics or machine acquisition settings. The local phase image features which provide more structural information about the local anatomy. Especially the consolidations are more dominantly represented and highly localized in the enhanced images.  

\begin{figure}
	\centering                                  
	\includegraphics[width=.9\linewidth]{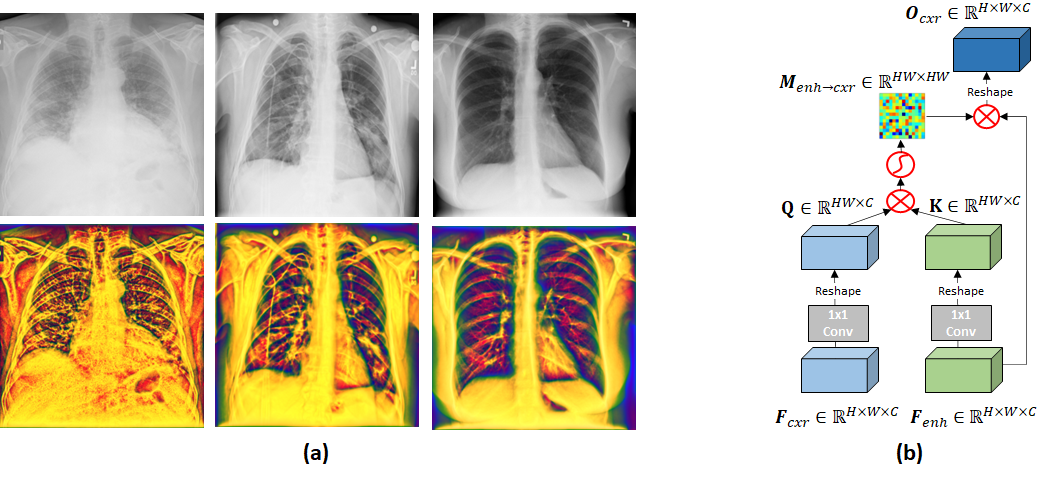}
	\caption{(a) The top row and bottom row are $CXR(x,y)$ and $MF(x,y)$ images respectively. The first column are from subjects who are diagnosed with COVID-19, the second column is from a subject who are diagnosed with pneumonia, and the last column is from a healthy subject. (b) An illustration of our Parallel-Attention(PA) block}
	\label{fig:enhance} 
\end{figure}

In the context of enhancing the appearance of CXR data, \cite{qisl2020} developed a local phase-based image enhancement method by designing bandpass quadrature filters. During this work, we deploy this image enhancement method for improving the representation of the CXR lung data (\cite{qisl2020}). The enhanced CXR image, termed multi-feature image $MF(x,y)$, is obtained by combining three different local phase image features: 1-Local weighted mean phase angle ($LwPA(x,y)$), 2- Weighted local phase energy ($LPE(x,y)$), and 3-Enhanced local energy attenuation image ($ELEA(x,y)$). $LPE(x,y)$ and $LwPA(x,y)$ image features are extracted by filtering the CXR image in frequency domain using monogenic filter and $\alpha$-scale space derivative (ASSD) bandpass quadrature filters (\cite{qisl2020}). $ELEA(x,y)$ image is extracted from the $LPE(x,y)$ image by modeling the scattering and attenuation effects of lung tissue inside a local region using L1 norm-based contextual regularization method (\cite{qisl2020}). We have used the filter parameters reported in \cite{qisl2020} for processing all the CXR images used in this work. Investigating Figure \ref{fig:enhance}, we can see that structural features inside the lung tissue are more dominant for COVID-19 CXR images compared to healthy lung tissue (last column Fig.\ref{fig:enhance}). The next section explains how the $MF(x,y)$ images are used to improve the diagnostic accuracy of our proposed deep learning model.

\subsection{Parallel-Attention Block}
Figure \ref{pipeline} (b) illustrates parallel-attention (PA) block for the original CXR images ($ CXR(x,y) $), and the same procedure is performed for the enhanced CXR images ($MF(x,y)$) by simply swapping the index \textit{cxr} and \textit{enh}. Two embedded feature maps $ \textbf{F}_{cxr} $, $\textbf{F}_{enh}$  \(\in \mathbb{R}^{H \times W \times C} \) from  $CXR(x,y)$ and $ MF(x,y) $ are fed to a PA block. In a PA block, $ \textbf{F}_{cxr} $ is fed to a 1$\times$1 convolution layer for feature adaption to produce a query feature map $ \textbf{Q} $  \(\in \mathbb{R}^{H \times W \times C} \), and meanwhile $ \textbf{F}_{enh} $ is fed to another 1 $\times$ 1 convolutional layer for feature adaption to produce a key feature map $ \textbf{K} $  \(\in \mathbb{R}^{H \times W \times C} \), as shown in Eq.\ref{eq:1}. \textbf{Q} and \textbf{K} are then reshaped to $ \mathbb{R}^{HW \times C} $. Then, matrix multiplication is performed between \textbf{Q} and \textbf{K} followed by a softmax layer to produce a attention matrix $ \textbf{M}_{enh {\scriptscriptstyle\rightarrow} cxr}$ \(\in \mathbb{R}^{HW \times HW} \), as shown in Eq.\ref{eq:2}. Next, $\textbf{F}_{enh}$ is reshaped to $\textbf{F}^{'}_{enh}$ \(\in \mathbb{R}^{H W \times C} \) and  multiples $ \textbf{M}_{enh {\scriptscriptstyle\rightarrow} cxr}$ to produce output feature map $\textbf{O}_{cxr}$  \(\in \mathbb{R}^{H \times W \times C} \), as shown in Eq.\ref{eq:3}. Through matrix multiplication between \textbf{Q} and \textbf{K}, our PA can effectively encode feature correlation between $ \textbf{F}_{cxr} $ and $ \textbf{F}_{enh} $ into attention map \textbf{M}, and thus two attention maps ( $ \textbf{M}_{enh {\scriptscriptstyle\rightarrow} cxr}$ and $ \textbf{M}_{cxr {\scriptscriptstyle\rightarrow} enh}$) are generated by our PA block. Correspondence can then be captured by multiplying its corresponding $ \textbf{F} $ as shown in Eq.\ref{eq:3}.

\begin{align} 
\textbf{Q} = \mathrm{conv}(\textbf{F}_{cxr}),  
\textbf{K} &= \mathrm{conv}(\textbf{F}_{enh}) \label{eq:1}\\ 
\textbf{M}_{enh {\scriptscriptstyle\rightarrow} cxr} &= \mathrm{softmax}(\textbf{QK}^T/\sqrt{C}) \label{eq:2}\\ 
\textbf{O}_{cxr} &= \textbf{M}_{enh {\scriptscriptstyle\rightarrow} cxr}\textbf{F}^{'}_{enh} \label{eq:3}
\end{align}

\subsection{Multi-Scale Parallel-Attention Fusion Network}
\begin{figure}
	\begin{center}
		\includegraphics[width=\linewidth]{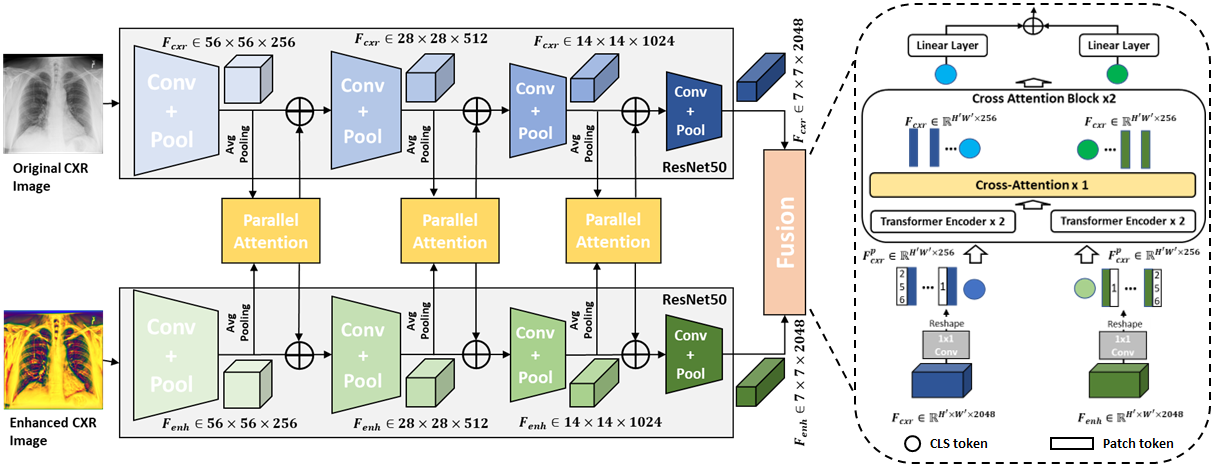}
		\caption{\textbf{Architecture}. An original CXR image and local-phase enhanced CXR image are fed to our proposed Res50-PA-ViT, which uses several PA blocks for the fusion of intermediate feature maps between two inputs. This fusion is applied at multiple resolutions (56$ \times $56, 28$ \times $28, 14$ \times $14) throughout the feature extractor. In the end, the outputs of two branches are fused together through cross-attention operation}
		\label{pipeline}
	\end{center}
\end{figure}

The network architecture of our propose multi-scale parallel-attention (MS-PA) fusion network is shown as Figure \ref{pipeline} (a). A $CXR(x,y)$ image and its corresponding $ MF(x,y) $ image as inputs are fed to two same CNN architectures, then several PA blocks are used for the fusion of intermediate feature maps from convolutional feature extractors between $CXR(x,y)$ and $ MF(x,y) $ images, and in the end a ViT with cross-attention is used for making the final decision. The convolutional feature extractors for the $CXR(x,y)$ and $ MF(x,y) $ inputs encode different aspects of the image at each scale. Therefore, we fuse these features at multiple scales throughout the convolutional encoder. 

As shown in Figure \ref{pipeline} (a), the intermediate feature maps $ \textbf{F}_{cxr}, \textbf{F}_{enh} $ \(\in \mathbb{R}^{H \times W \times C} \) are fed to a PA block, which produces two outputs $ \textbf{O}_{cxr} $, $\textbf{O}_{enh}$  \(\in \mathbb{R}^{H \times W \times C} \). The each output is then fed back into each of the individual branch using an element-wise summation with the existing feature maps. The mechanism described above constitutes feature fusion at a single scale. This fusion is applied multiple times throughout the convolutional feature extractors of the $CXR(x,y)$ and $ MF(x,y) $ branches at different resolutions. However, processing feature maps at high spatial resolutions is computationally expensive, specially when your inputs have a high resolution. Therefore, we downsample higher resolution feature maps from the early convolutional feature extractors using average pooling (H$ = $W$ = $7) to the resolution of feature map from the last convolutional feature extractor before passing them as inputs to a PA block and upsample the output to the original resolution using bilinear interpolation before element-wise summation with the existing feature maps. 

After PA fusion at multiple scales, feature maps $ \textbf{F}_{cxr}, \textbf{F}_{enh} $ \(\in \mathbb{R}^{H^{'} \times W^{'} \times C^{'}} \) are obtained at the last convolutional feature extractors from the $CXR(x,y)$ and $ MF(x,y) $ branches. Then, we projected encoded feature maps $ \textbf{F}_{cxr} $ and $ \textbf{F}_{enh} $ of dimension $ C^{'} $ into $ \textbf{F}_{cxr}^{p} $ and $ \textbf{F}_{enh}^{p} $  with dimension of 256 using 1 $\times$ 1 convolution kernel. Then $ \textbf{F}_{cxr}^{p} $ and $ \textbf{F}_{enh}^{p} $ are reshaped to $ \mathbb{R}^{H^{'} W^{'} \times 256} $. We prepended additional CLS tokens and added positional embedding for incorporating global context to $ \textbf{F}_{cxr}^{p} $ and $ \textbf{F}_{enh}^{p} $. Two cross-attention blocks are applied to further exchange information between $CXR(x,y)$ and $ MF(x,y)$ (\cite{mf_vit}). As $ \textbf{F}_{cxr}$ and $\textbf{F}_{enh} $ already encodes the representations for important findings from $CXR(x,y)$ and $ MF(x,y) $, we adopted relatively simple architecture for Transformer encoder with two encoder layers with four attention heads. In the end, CLS tokens from two branches are projected into two 3-dimensional feature vectors and summed together.


\section{Experiments}

\subsection{Datasets and Partitioning.}
In this study, all images were collected from six public data repositories, which are BIMCV (\cite{bimcv}), COVIDx (\cite{Wang2020_covid}), COVID-19-AR (\cite{Rural}), MIDRC-RICORD-1c (\cite{RSNA}), COVID-19-IR (\cite{Germany}) and COVID-19-NY-SBU (\cite{SBU}). Total dataset includes three classes: normal, pneumonia, and COVID-19. The numbers of images for each class are summarized in Table \ref{dataset_img}. COVID-19 CXR images from both (BIMCV\cite{bimcv}) and COVIDx (\cite{Wang2020_covid}) datasets plus 2567 randomly selected CXR images from normal and pneumonia classes composed the 'Evaluation Dataset'. To evaluate the generalization ability of models in various institutional settings, we used the rest normal and pneumonia CXR images from \cite{Wang2020_covid} and four COVID-19 datasets collected from different institutions with different devices as 'Test Dataset-2'.

\begin{table}
	\caption{Summary of data distribution}
	\label{dataset_img}
	\centering
	\scalebox{0.8}{
		\begin{tabular}{l|c|c|c|c|c|c|c|c} 
			\hline
			\multirow{2}{*}{\textbf{Class}}&\multirow{2}{*}{\textbf{Total}} & \multicolumn{2}{c|}{\textbf{Evaluation Dataset}} & \multicolumn{5}{c}{\textbf{Test Dataset-2}}\\
			\cline{3-9}
			&         & COVIDx & BIMCV & COVIDx & AR & 1c & IR & SBU \\ 
			\hline
			Normal      & 8,851   & 2,567     & -   & 6,284 & -   & -   & - & - \\ 
			\hline
			Pneumonia   &  6,045  & 2,567     & -   & 3,478 & -   & -   & - & -\\                            
			\hline
			COVID-19    & 10,776  &   400   & 2,167 & -    & 249 & 977 & 243 & 6,740 \\ 
			\hline
			Total image & 25,672  &  5,534  & 2167 & 9,762 & 249 & 977 & 243 & 6,740 \\ 
			\hline
	\end{tabular}}
\end{table}

%
%

\subsection{Implementation Details}
We choose ResNet-50 pretrained on ImageNet that has the best performance reported by \cite{qisl2020} and DenseNet-121 with PCAM pooling pretrained on CheXpert (\cite{chexpert}) that achieved the first place in CheXpert challenge in 2019 (\cite{pcam}) as the backbone network architectures. CheXpert (\cite{chexpert}) dataset has 223,415 CXR images containing 13 labeled observations including no finding, enlarged cardiomediastinum, cardiomegaly, lung opacity, lung lesion, edema, consolidation, pneumonia, atelectasis, pneumothorax, pleural effusion, pleural other, and support devices. Binary cross-entropy (BCE) losses were used as the loss function. For training of the proposed model for COVID-19 diagnosis, we used SGD optimizer (momentum = 0.9) with learning rate of 0.001. The model was trained for 35 epochs with cosine warm-up scheduler (warm-up epochs = 4), with batch size set to 32. Five-fold cross-validation was performed on 'Evaluation Dataset' for training and testing the proposed methods. Our augmentation method includes random translation, rotation, changing lighting condition, normalization, horizontal flip and resizing. We adopted mean overall accuracy (Acc) as our primary evaluation metric, but also calculated sensitivity, precision, F-1 score and area under the ROC curve (AUC). All experiments were performed with Python version 3.6 and PyTorch 1.10 on two Nvidia 1080Ti.


\section{Results}
\textbf{Diagnostic performance of proposed methods: }\noindent
The diagnostic performances of our models are provided in Table \ref{PA-ViT}. Dense121-PA-ViT achieves the average AUCs of 0.990, 0.992, average precision 0.960, 0.973, average sensitivity 0.963, 0.970, average F-1 score 0.963, 0.973, and average accuracy of 96.65\%, 97.88\% in the Evaluation dataset and Test dataset-2. In the Evaluation Dataset, both Res50-PA-ViT and Dense121-PA-ViT show 0.99 of precision, sensitivity, F-1 score for COVID-19 class, and Res50-PA-ViT achieves 99.12\% accuracy for COVID-19 class slightly better than Dense121-PA-ViT. In the Test Dataset-2, Dense121-PA-ViT has the highest precision, sensitivity, F-1 score, AUC, and accuracy for all classes compared to Res50-PA-ViT. 

\begin{table}
	\caption{Diagnostic performance of proposed methods on two datasets. Top table is results of Res50-PA-ViT and bottom table is results of Dense121-PA-ViT}
	\label{PA-ViT}
	\centering 
	\scalebox{0.8}{
		\begin{tabular}{l|c|c|c|c|c|c|c|c} 
			\hline
			\multirow{2}{*}{\textbf{Metrics}} & \multicolumn{4}{c|}{\textbf{Evaluation Dataset}} & \multicolumn{4}{c}{\textbf{Test Dataset-2}} \\
			\cline{2-9}
			& Avg. & Normal & Pneumonia & COVID & Avg. & Normal & Pneumonia & COVID \\ 
			\hline
			AUC        &  0.990 & 0.989 & 0.983 & 0.998 & 0.991 & 0.992 & 0.983 & 0.999 \\ 
			\hline
			Precision  &  0.957 & 0.94  &  0.94 & 0.99  & 0.957  & 0.96  & 0.92 & 0.99 \\ 
			\hline
			Sensitivity&  0.953  & 0.94  &  0.93 & 0.99  & 0.960  & 0.96  & 0.93 & 0.99 \\                            
			\hline
			F-1 Score  & 0.957  & 0.94  &  0.94 & 0.99  & 0.960 & 0.96 & 0.93 & 0.99 \\ 
			\hline
			Accuracy(\%)   & 96.04  & 94.66 & 94.34 & 99.12 & 97.05 & 96.54 & 95.43 & 99.19 \\
			\hline
			\hline
			\multirow{2}{*}{\textbf{Metrics}} & \multicolumn{4}{c|}{\textbf{Evaluation Dataset}} & \multicolumn{4}{c}{\textbf{Test Dataset-2}} \\
			\cline{2-9}
			& Avg. & Normal & Pneumonia & COVID & Avg. & Normal & Pneumonia & COVID \\ 
			\hline
			AUC           & 0.990 & 0.987 & 0.985 & 0.998 & 0.992 & 0.993 & 0.985 & 0.999 \\  
			\hline
			Precision     & 0.960 &   0.94  & 0.95 & 0.99 & 0.973  & 0.96 & 0.96 & 1.00 \\ 
			\hline
			Sensitivity   & 0.963 &   0.96  & 0.94 & 0.99 & 0.970  & 0.97 & 0.94 & 1.00\\                            
			\hline
			F-1 Score    & 0.963 &   0.95  & 0.95 & 0.99 &  0.973 & 0.97 & 0.95 & 1.00 \\ 
			\hline
			Accuracy(\%)     & 96.56  & 95.43 & 95.22 & 99.03 & 97.88 & 97.01 & 96.84 & 99.80 \\
			\hline
	\end{tabular}}
\end{table}

\noindent\textbf{Comparison with baseline and SOTA models: }
Table \ref{Compare} shows the diagnostic performances of our proposed models compared to the baseline and SOTA models. Res50-CXR and Res50-Enh were used as baseline models (\cite{qisl2020}). Dense121-ViT, Res50-mid, Res50-late fusion architectures proposed by \cite{chxpert_vit} and \cite{qisl2020} respectively were used as SOTA models. Dense121-ViT has the same backbone structure as our proposed Dense121-PA-ViT, which was pre-trained using Chexpert (\cite{chexpert}) with PCAM pooling (\cite{pcam}) and finetuned using Evaluation Dataset. As shown in Table \ref{Compare}, our proposed Dense121-PA-ViT obtains the highest overall accuracy in both datasets and achieves statistically significant improvement compared with the best model in baseline and SOTA methods with paired t-test ($ p<0.05 $). Our proposed Dense121-PA-ViT achieves significantly higher accuracy in identification of pneumonia cases compared to Dense121-ViT-Enh, the best SOTA model shown as in Table \ref{Compare}. Test Dataset-2 incorporates more COVID-19 CXR images, which were collected from multiple organizations with different devices and settings. Investigating Table \ref{Compare}, we can see that the performances of models drop significantly when testing on ‘Test Dataset-2’ if the training was performed using $CXR(x,y)$ only. In comparison when the training data involved $MF(x,y)$ as a secondary dataset the performance and the generalization ability of the models improved. We also compared the performance of our proposed models with baseline and SOTA models in terms of AUC. Table \ref{auc_compare} in the Appendix \ref{appdenix:b} shows the comparison results. Furthermore, an additional ultrasound (US) dataset was used to demonstrate the proposed method can be generalized to classification task from other image modalities. Similarly, the proposed Dense121-PA-ViT achieves the best performance in comparison with baseline and SOTA models. The detailed results are presented in the Appendix \ref{appendix:c}. 

\begin{table}
	\caption{Comparison with baseline and SOTA methods in accuracy (\%)}
	\label{Compare}
	\centering 
	\scalebox{0.80}{
		\begin{tabular}{l|c|c|c|c|c|c|c|c} 
			\hline
			\multirow{2}{*}{\textbf{Methods}} &\multicolumn{4}{c|}{\textbf{Evaluation Dataset}} & \multicolumn{4}{c}{\textbf{Test Dataset-2}} \\
			\cline{2-9}
			& Avg. & Normal & Pneumonia & COVID & Avg. & Normal & Pneumonia & COVID \\ 
			\hline
			Res50-CXR  &  94.20 &  93.79  & 90.44 & 98.36 &  88.12 &  91.77  & 87.79 & 86.72 \\ 
			\hline
			Res50-Enh&   94.13 & 94.43 & 91.12 & 96.83 & 95.61 & 94.72 & 90.71 & 98.37 \\ 
			\hline \hline                        
			Res50-mid   &    \multirow{2}{*}{94.88}  &  \multirow{2}{*}{94.08}  & \multirow{2}{*}{91.75} & \multirow{2}{*}{98.83} &  \multirow{2}{*}{95.83} & \multirow{2}{*}{94.70} & \multirow{2}{*}{90.87} & \multirow{2}{*}{97.40} \\
			(\cite{qisl2020}) &           &         &       &     &       &      &       &       \\
			\hline
			Res50-late    &  \multirow{2}{*}{95.36} & \multirow{2}{*}{95.67} & \multirow{2}{*}{91.52}  & \multirow{2}{*}{98.90} & \multirow{2}{*}{95.12} & \multirow{2}{*}{95.95} & \multirow{2}{*}{90.56} & \multirow{2}{*}{95.80} \\
			(\cite{qisl2020})    &   &  &   &  &  &  &  &  \\
			\hline
			Dense121-ViT-CXR &  \multirow{2}{*}{93.74} & \multirow{2}{*}{91.86} &  \multirow{2}{*}{90.62} &  \multirow{2}{*}{98.75} &  \multirow{2}{*}{90.14} & \multirow{2}{*}{90.73} & \multirow{2}{*}{88.29} & \multirow{2}{*}{90.32} \\
			(\cite{chxpert_vit}) &   &  &   &   &   &  &  &  \\  
			\hline
			
			Dense121-ViT-Enh &  94.87   &  \textbf{96.69 }& 90.54 & 97.39 & 96.77 & 96.09 & 90.68 & \textbf{99.88 } \\                            
			\hline\hline
			Res50-PA-ViT(ours)  & 96.04  & 94.66 & 94.34 & \textbf{99.12} & 97.05 & 96.54 & 95.43 & 99.19 \\	
			\hline
			Dense121-PA-ViT(ours) & \textbf{96.56 } & 95.43 & \textbf{95.22 }& 99.03 & \textbf{97.88} & \textbf{97.01} & \textbf{96.84} & 99.80 \\ 	
			\hline
	\end{tabular}}
\end{table}

\noindent\textbf{Ablation Study: }We conducted an ablation study to investigate the effect of PA block for training a multi-feature network. The results of Table \ref{ablation} suggest that adding PA block can further improve performance in two types of multi-feature fusion models, Res50-late and Res50-mid, reported in \cite{qisl2020}. PA blocks were added to the fore-mentioned models, denoted as Res50-PA-late and Res50-PA-mid. The network architecture of Res50-PA-mid (Figure \ref{fig:res50_pa_mid}) and Res50-PA-late (Figure \ref{fig:res50_pa_late}) are shown in the Appendix \ref{appendix:a}. In the Evaluation Dataset, Res50-PA-mid achieves statistically significant improvement compared to Res50-mid without using PA blocks with paired t-test ($ p<0.05 $). In the Test Dataset-2, Res50-PA-late is statistically significantly better than Res50-late with paired t-test($ p<0.05 $). 
\begin{table}
	\caption{Performance of multi-feature fusion model with and without PA blocks}
	\label{ablation}
	\centering 
	\scalebox{0.8}{
		\begin{tabular}{l|c|c|c|c|c|c|c|c} 
			\hline
			\multirow{2}{*}{\textbf{Methods}} &\multicolumn{4}{c|}{\textbf{Evaluation Dataset}} & \multicolumn{4}{c}{\textbf{Test Dataset-2}} \\
			\cline{2-9}
			& Avg. & Normal & Pneumonia & COVID & Avg. & Normal & Pneumonia & COVID \\
			\hline
			Res50-mid   &    94.88  &  94.08  & 91.75 & 98.83 &  95.83 & 94.70 & 90.87 & 97.40 \\ 
			\hline
			Res50-late    &  95.36 & 95.67 & 91.52  & 98.90 & 95.12 & 95.95 & 90.56 & 95.80 \\
			\hline\hline
			Res50-PA-mid   &  95.94  & 94.36  & 94.23 & 99.22 & 96.28 & 95.58 & 94.95 & 98.30 \\ 
			\hline
			Res50-PA-late   & 95.87  & 94.44 & 93.96 & 99.22 & 96.54 & 95.45 & 95.02 & 99.16 \\	
			\hline
			
	\end{tabular}}
\end{table}

\begin{figure}
	\centering                                  
	\includegraphics[width=0.9\linewidth]{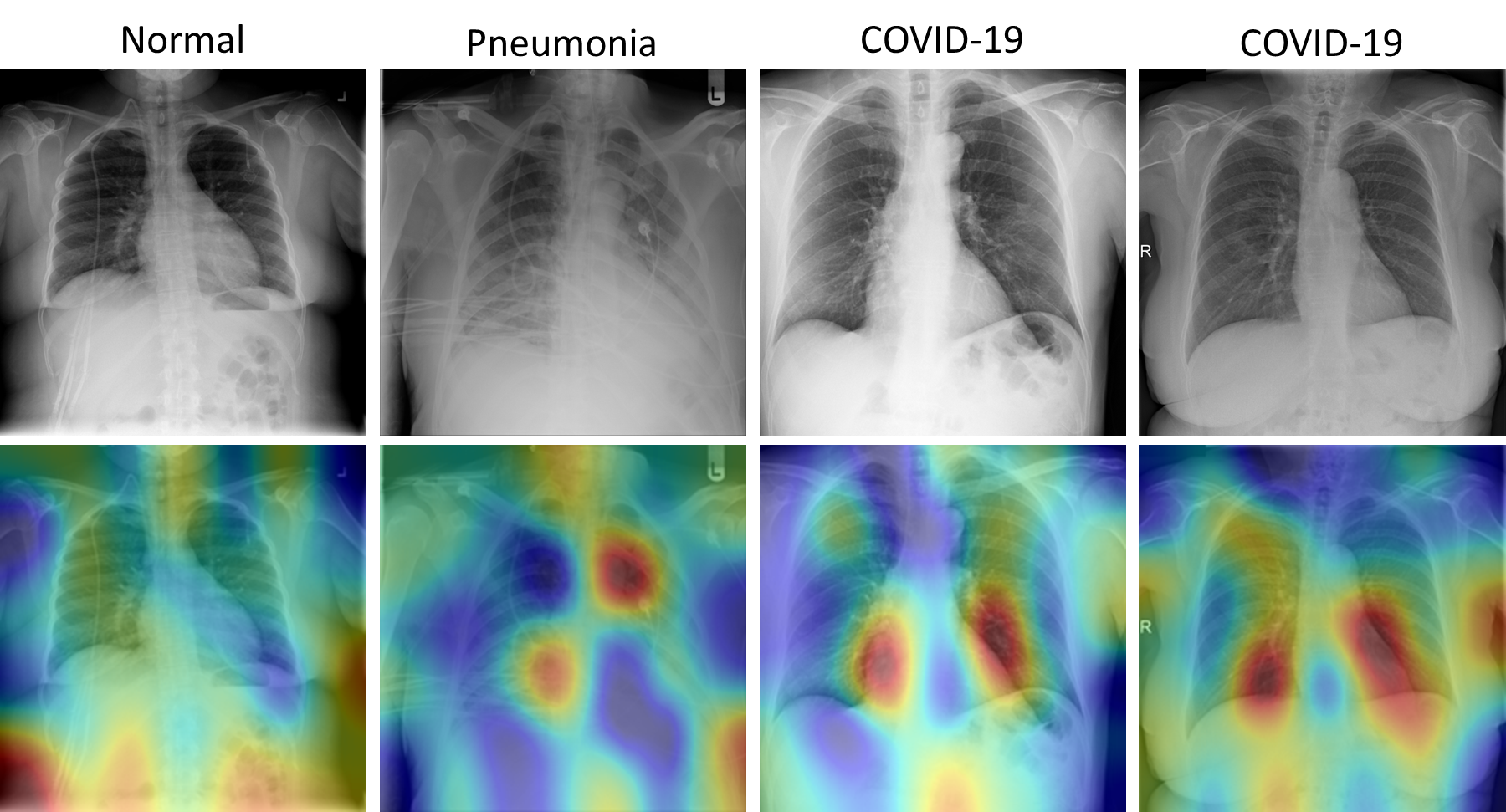}
	\caption{Grad-CAM visualization for each class.}
	\label{fig:grad_cam} 
\end{figure}

\noindent\textbf{Grad-CAM Visualization Results: }Figure \ref{fig:grad_cam} shows the examples of Grad-CAM Map (\cite{grad_cam}) based visual representations for each disease classes. The Grad CAM Map visualization has the capability to highlight the regions in the lung that are significant for disease predictions as well as disease development.

The first column of Figure \ref{fig:grad_cam} shows a normal case having highlighted regions surrounding the lungs. The Grad-CAM based heatmap is similar to the reviewing process of a radiologist. During the reviewing, because the lungs are clear, the radiologist's eye gaze skips around the image without focus on the lung (\cite{karargyris2021creation}). The second column of Figure \ref{fig:grad_cam} shows pneumonia patient's lungs with affected regions highlighted in dark red which indicate severe infection. It is similar to the physician's eye gaze focuses on the focal lung opacity. The last two columns of Figure \ref{fig:grad_cam} show multi-focal infected areas in the lungs, which is common for the COVID-19 patients. The figure clearly shows that our proposed method recognizes and differentiates relevant impacted areas from COVID-19 and other pneumonia images.  

To have a better understanding of the model's misprediction, we exemplified the failure cases by the proposed model as shown in Figure \ref{fig:grad_cam}. Its confusion could be explained by the generated Grad-CAM visual interpretations. Figure \ref{fig:grad_cam}, (a) shows the model misclassified a case of pneumonia as COVID-19, as the location and distribution of the consolidations resemble those of COVID-19 (multi-focal consolidations), and (b) shows a severe COVID-19 case that was confused as pneumonia, where a focal cavitary lesion is similar to pneumonia case caused by bacteria infection. Our proposed Dense121-PA-ViT achieves a significantly higher accuracy compared to best SOTA model in identification of pneumonia cases (95.22\% vs. 91.52\% in the Evaluation Dataset; 96.84\% vs. 90.87\% in the Test Dataset-2).

\begin{figure}
	\centering                                  
	\includegraphics[width=0.9\linewidth]{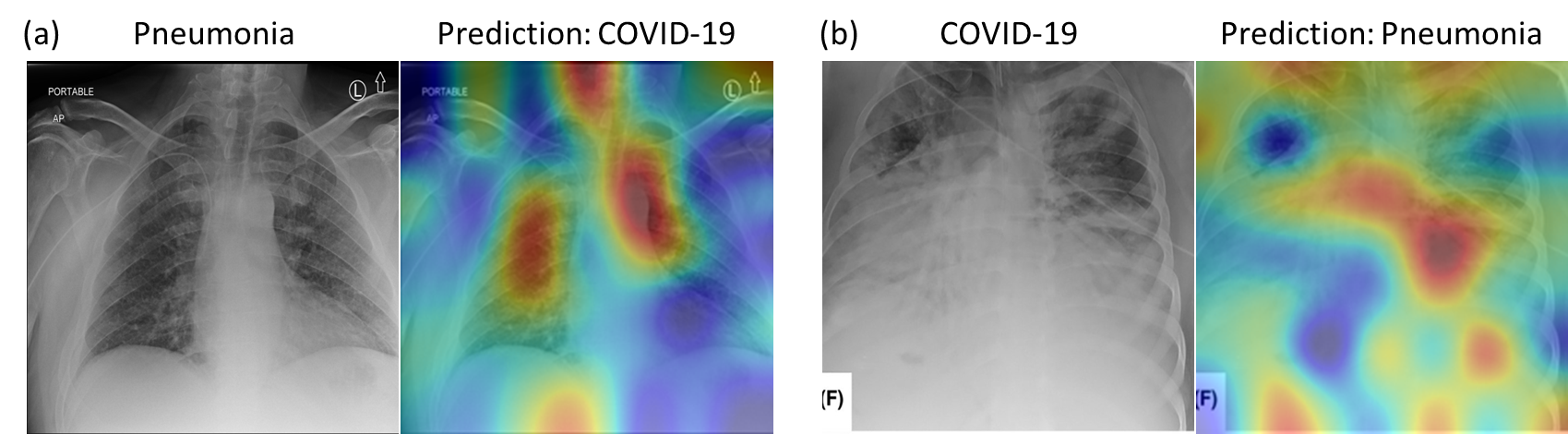}
	\caption{Examples of failure cases of the proposed model.}
	\label{fig:mis} 
\end{figure}

\section{Conclusion}
In this work, we proposed a novel multi-feature fusion model for COVID-19 CXR diagnosis by applying parallel-attention block at multiple scales. The novelty of this study lies in leveraging self-attention mechanism to extract information from both original CXR images and local-phase enhanced CXR images at multiple scales. In addition, the local-phase based image enhancement method can eliminate intensity variance caused by different device setting and subjects characteristics so that increasing the model generality and stability. The experimental results on various COVID-19 test datasets confirm that our model not only achieves SOTA performance in the diagnosis of COVID-19 and other infectious disease but also retains stable performance irrespective of the external settings, which is a sine-qua-non for widespread application of system. Future work will include the extension of the method for accurately detecting the presence of multiple diseases from CXRs.

\ethics{The work follows appropriate ethical standards in conducting research and writing the manuscript, following all applicable laws and regulations regarding treatment of animals or human subjects.}

\coi{There is no conflict of interest. }

\newpage
\bibliography{sample}


\newpage
\appendix
%
%
\section{Network Architectures}
\label{appendix:a}
\subsection{Architecture of Res50-PA-mid:}
\begin{figure}
	\centering                                  
	\includegraphics[width=0.9\linewidth]{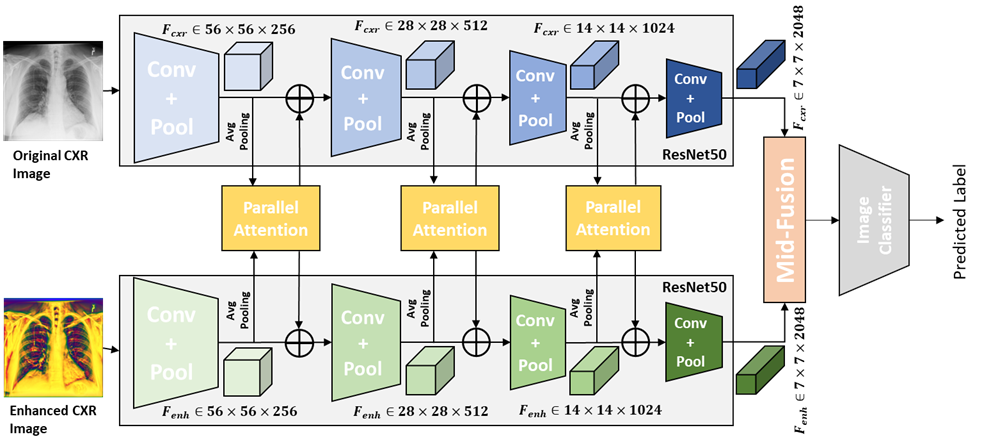}
	\caption{Network architecture of Res50-PA-mid}
	\label{fig:res50_pa_mid} 
\end{figure}

Res50-PA-mid, shown as Figure \ref{fig:res50_pa_mid}, has the same feature extractor as the Res50-PA-ViT, but it uses middle fusion with Convolutional operation reported in \cite{qisl2020} instead. 

\subsection{Architecture of Res50-PA-late:}
\begin{figure}
	\centering                                  
	\includegraphics[width=0.9\linewidth]{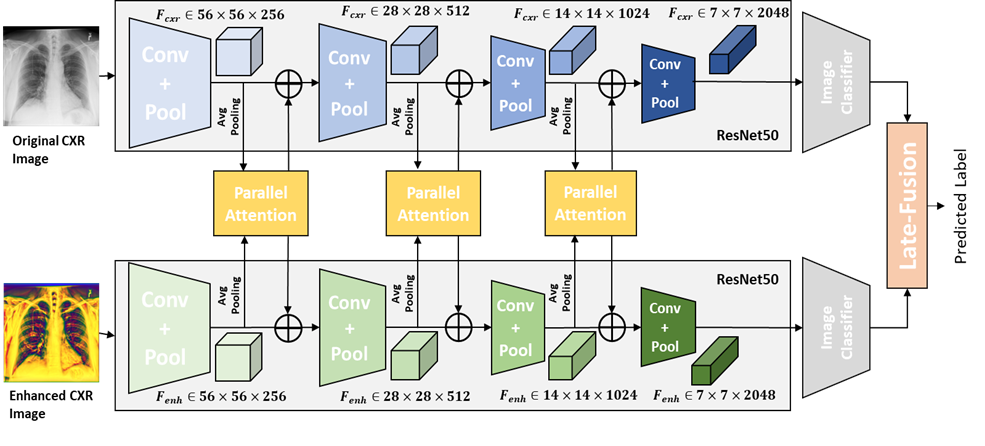}
	\caption{Network architecture of Res50-PA-late}
	\label{fig:res50_pa_late} 
\end{figure}

Res50-PA-late, shown as Figure \ref{fig:res50_pa_late}, also has the same feature extractor as the Res50-PA-ViT. After feature extraction, the features are feed into image classifier, which has the same structure of ResNet50. The final decision is made based on the joint correspondence using sum operation reported in \cite{qisl2020}. 

\section{Performance of Proposed Methods in terms of AUC}
\label{appdenix:b}
\subsection{Comparison with baseline and SOTA models:}
To reinforce the performance of proposed methods in the detection of COVID-19, we also report the results of AUC tested on Evaluation Dataset and Test Dataset-2 in comparison with baseline and SOTA models shown as Table \ref{auc_compare}. Dense121-PA-ViT achieves the highest average AUCs compared to other methods testing on two datasets. Res50-PA-mid and Res50-PA-late show a better performance compared to Res50-mid and Res50-late, two multi-feature fusion models without PA blocks reported in \cite{qisl2020}. It demonstrates using PA blocks at multi-scales for feature fusion can further improve the model performance.    

\begin{table}
	\caption{Comparison with baseline and SOTA models in terms of AUC}
	\label{auc_compare}
	\centering 
	\scalebox{0.8}{
		\begin{tabular}{l|c|c|c|c|c|c|c|c} 
			\hline
			\multirow{2}{*}{\textbf{Methods}} &\multicolumn{4}{c|}{\textbf{Evaluation Dataset}} & \multicolumn{4}{c}{\textbf{Test Dataset-2}} \\
			\cline{2-9}
			& Avg. & Normal & Pneumonia & COVID & Avg. & Normal & Pneumonia & COVID \\
			\hline
			Res50-CXR   & 0.983 & 0.978 & 0.973 & 0.998 & 0.973 & 0.987 & 0.949 & 0.982  \\ 
			\hline
			Res50-Enh    & 0.984 & 0.983 & 0.973  & 0.998 & 0.985 & 0.989 & 0.968 & 0.998 \\
			\hline
			Res50-mid     & 0.986  & 0.984 & 0.977 & 0.998 & 0.985 & 0.989 & 0.967 & 0.999 \\ 
			\hline
			Res50-late    & 0.986 & 0.983 & 0.977 & 0.998 & 0.984 & 0.992 & 0.963 & 0.999 \\
			\hline
			Res50-PA-mid     & 0.988 & 0.987 & 0.979 & 0.999 & 0.987 & 0.994 & 0.968 & 0.999 \\ 
			\hline
			Res50-PA-late    & 0.988 & 0.985 & 0.979 & \textbf{0.999} & 0.988 & \textbf{0.994} & 0.973 & 0.999 \\
			\hline
			Dense121-ViT-CXR    & 0.984 & 0.981 & 0.977 & 0.996 & 0.976 & 0.990 & 0.955 & 0.983 \\ 
			\hline
			Dense121-ViT-Enh    & 0.986 & 0.982 & 0.977 & 0.998 & 0.988 & 0.992 & 0.974 & 0.999 \\
			\hline\hline
			Res50-PA-ViT(ours)   & 0.990 & \textbf{0.989} & 0.983 & 0.998 & 0.991 & 0.992 & 0.983 & 0.999 \\ 
			\hline
			Dense121-PA-ViT(ours)   & \textbf{0.990} & 0.987 & \textbf{0.985} & 0.998 & \textbf{0.992} & 0.993 & \textbf{0.985} & \textbf{0.999} \\	
			\hline			
	\end{tabular}}
\end{table}

\section{Evaluation on Ultrasound (US) Images}
\label{appendix:c}
\subsection{Dataset}
\begin{figure}
	\centering                                  
	\includegraphics[width=0.6\linewidth]{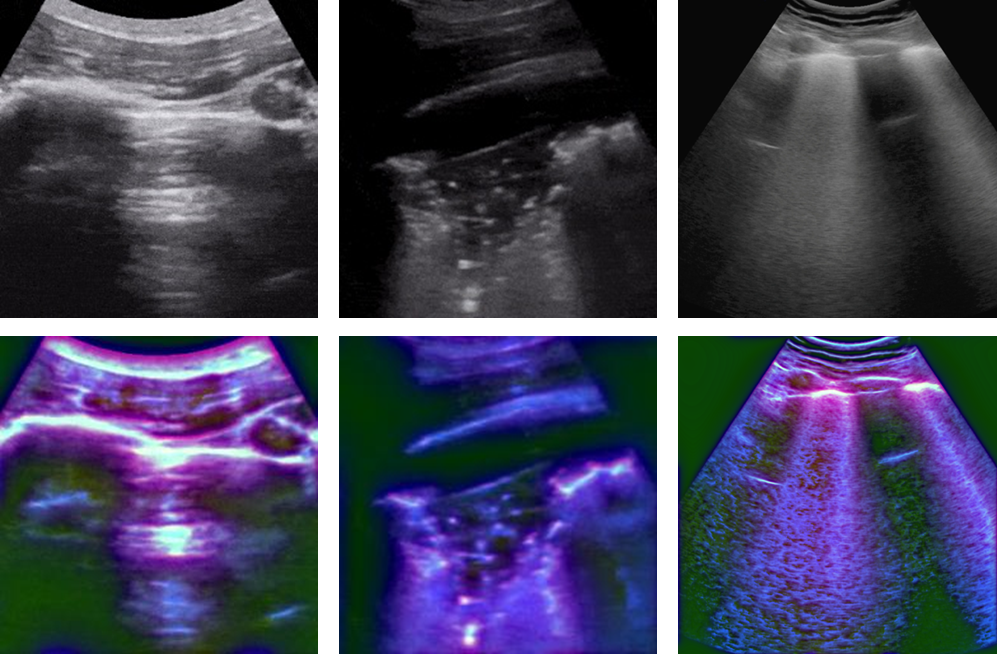}
	\caption{The top row and bottom row are $US(x,y)$ and $MF(x,y)$ images respectively. From left to right are normal, pneumonia, and COVID-19 cases.}
	\label{fig:us_enhance} 
\end{figure}

An publicly available US dataset (\cite{us_dataset}) was choose as an additional dataset. The dataset has 2825 US images, including 852 COVID-19, 707 bacterial pneumonia, and 1266 healthy images, which are extracted from 157 convex ultrasound videos at a frame rate of 3Hz with maximal 30 frames per video and includes 56 convex ultrasound static images. All these images are split into five folds on a video level since the consecutive US frames are highly correlated. The reported local phase-based image enhancement method was applied to the US images to obtain the multi-feature $MF(x,y)$ images. Figure \ref{fig:us_enhance} shows the original US images and the corresponding enhanced US images for each class.

\subsection{Results}
Table \ref{US_PA_ViT} shows the diagnostic performance of our proposed method tested on the US dataset. Dense121-PA-ViT achieves the average AUC of 0.944, average precision 0.880, average sensitivity 0.877, average F-1 score 0.877, and average accuracy of 89.52\%. Dense121-PA-ViT has the highest precision, sensitivity, F-1 score, AUC, and accuracy for all classes compared to Res50-PA-ViT.

Table \ref{us_acc_auc_compare} presents the diagnostic performances of our proposed models compared to the baseline and SOTA models in average accuracy and AUC. As shown in Table \ref{us_acc_auc_compare}, Dense121-PA-ViT obtains the highest overall accuracy and class accuracy for all classes. And the proposed Dense121-PA-ViT achieves statistically significant improvement compared with the best baseline and SOTA methods with paired t-test ($ p<0.05 $) in overall accuracy and class accuracy. Investigating Table \ref{us_acc_auc_compare}, we can also see the Dense121-PA-ViT obtains the highest overall AUC and class AUC except normal class. The Dense121-ViT-Enh achieves the highest AUC in identification of normal class.

\begin{table}
	\caption{Diagnostic performance of proposed methods evaluated on the US dataset}
	\label{US_PA_ViT}
	\centering 
	\scalebox{0.8}{
		\begin{tabular}{l|c|c|c|c|c|c|c|c} 
			\hline
			\multirow{2}{*}{\textbf{Metrics}} & \multicolumn{4}{c|}{\textbf{Res50-PA-ViT (Ours)}} & \multicolumn{4}{c}{\textbf{Dense121-PA-ViT (Ours)}} \\
			\cline{2-9}
			& Avg. & Normal & Pneumonia & COVID & Avg. & Normal & Pneumonia & COVID \\ 
			\hline
			AUC        &  0.916 & 0.914 & 0.915	& 0.919	& 0.944	& 0.943	& 0.936 & 0.951 \\ 
			\hline
			Precision  & 0.853 & 0.87 &	0.85 & 0.84	& 0.880 & 0.88 & 0.87 & 0.89 \\ 
			\hline
			Sensitivity& 0.847 & 0.87 &	0.86 & 0.81 & 0.877 & 0.89 & 0.89 &	0.85 \\                            
			\hline
			F-1 Score  & 0.850	& 0.87	& 0.85 & 0.83 &	0.877 &	0.88 & 0.88 & 0.87 \\ 
			\hline
			Accuracy(\%)   & 87.42 & 87.36 & 88.99 & 85.91 & 89.52 & 88.53 & 90.83 & 89.20 \\
			\hline
	\end{tabular}}
\end{table}

\begin{table}
	\caption{Comparison with baseline and SOTA models in accuracy (\%) and AUC}
	\label{us_acc_auc_compare}
	\centering 
	\scalebox{0.8}{
		\begin{tabular}{l|c|c|c|c|c|c|c|c} 
			\hline
			\multirow{2}{*}{\textbf{Methods}} &\multicolumn{4}{c|}{\textbf{Accuracy for US Dataset}} & \multicolumn{4}{c}{\textbf{AUC for US Dataset}} \\
			\cline{2-9}
			& Avg. & Normal & Pneumonia & COVID & Avg. & Normal & Pneumonia & COVID \\
			\hline
			Res50-CXR   &82.55&	86.63&	81.38&	76.33 & 0.870&	0.902&	0.867&	0.840  \\ 
			\hline
			Res50-Enh    &83.71&88.02&	85.41&	73.72& 0.878&	0.931&	0.884&	0.820 \\
			\hline
			Res50-mid     &83.04&	86.54&	85.11&	76.13 & 0.876&	0.904&	0.884&	0.841 \\ 
			\hline
			Res50-late    & 84.76&	89.11&	85.61&	77.18& 0.893&	0.941&	0.897&	0.842 \\
			\hline
			Dense121-ViT-CXR    & 86.34&	86.54&	88.21&	84.27 & 0.907&	0.920&	0.935&	0.868 \\ 
			\hline
			Dense121-ViT-Enh    & 87.06&	88.18&	87.41&	85.60 & 0.913&	\textbf{0.949}&	0.928&	0.862 \\
			\hline\hline
			Res50-PA-ViT(ours)   & 87.42 & 87.36 & 88.99 & 85.91  & 0.916&	0.914&	0.915&	0.919 \\ 
			\hline
			Dense121-PA-ViT(ours)   & \textbf{89.52} & \textbf{88.53} & \textbf{90.83} & \textbf{89.20}  & \textbf{0.944}&	0.943&	\textbf{0.936}&	\textbf{0.951} \\	
			\hline			
	\end{tabular}}
\end{table}

\end{document}